\documentclass[10pt]{article}
\usepackage{graphicx}
\usepackage{dcolumn}
\usepackage{bm}
\usepackage{amsmath}
\usepackage{amssymb}
\usepackage{color}
\usepackage{natbib}
\bibpunct{(}{)}{;}{a}{,}{,}
\newcommand{\be}{\begin{eqnarray}}
\newcommand{\en}{\end{eqnarray}}
\newcommand{\ben}{\begin{eqnarray*}}
\newcommand{\enn}{\end{eqnarray*}}

\newcommand{\bi}{\begin{itemize}}
\newcommand{\ei}{\end{itemize}}

\newcommand{\hv}{\fontfamily{phv}\selectfont}

\title{{\bf Non-affine fields in solid-solid transformations: the structure and stability of a product droplet}}

\author{Arya Paul, \\ 
Advanced Materials Research Unit, S.N.Bose National Centre for Basic Sciences, \\ JD Block, Sector-III, Salt Lake, Kolkata - 700098, India. \\ \\
Surajit Sengupta, \\
Centre for Advanced Materials, Indian Association for the Cultivation of Science, \\ 2A \& 2B Raja S.C. Mullick Road, Jadavpur, Kolkata - 700032, India. \\ \\
Madan Rao,\\
 Raman Research Institute, C. V. Raman Avenue, Sadashivanagar, \\ Bangalore - 560080, India. \\
and \\
 National Centre for Biological Sciences, Tata Institute of Fundamental Research, \\ GKVK, Bellary Road, Bangalore 560065, India.
}

\date{\today}
\begin{document}
\maketitle
\abstract{
We describe the microstructure, shape and dynamics of growth of a droplet of martensite nucleating in a 
parent austenite during a solid-solid transformation, using a Landau theory written in terms of  conventional affine, elastic deformations and  {\em non-affine} degrees of freedom. Non-affineness, $\phi$, serves as a source of strain incompatibility and screens long-ranged elastic interactions. It is produced wherever the local stress exceeds a threshold and anneals diffusively thereafter. A description in terms of $\phi$ is inevitable when the separation between defect pairs, possibly generated during the course of the transformation, is small. 
Using a variational calculation, we find three types of stable solutions ({\hv I}, {\hv II} and {\hv III}) for the structure of the product droplet depending on the scaled mobilities of $\phi$ parallel and perpendicular to the parent-product interface and the stress threshold. In {\hv I}, $\phi$ is vanishingly small, {\hv  II} involves 
large $\phi$ localized in regions of high stress within the parent-product interface and  {\hv  III} where $\phi$ completely wets the parent-product interface. While width $l$ and size $W$ of the twins follows $l\propto\sqrt{W}$ in solution {\hv I}, this relation does not hold for {\hv II} or {\hv III}. We obtain a dynamical phase diagram featuring these solutions and argue that they represent specific microstructures such as twinned or dislocated martensites.  
}

\section{Introduction}
%

Unlike the problem of solidification\, \,\citep{cahn}, where a crystal nucleates and grows from a melt, nucleation of a solid within another solid may encounter specific complications arising from crystallographic incompatibility of the parent and product solids. In most cases it is geometrically impossible to fit a nucleus of the homogeneous product within the parent (austenite) matrix without gaps. To maintain continuity therefore, the system needs to adapt, which it does using several viable strategies. A common possibility is the simultaneous nucleation of several crystallographically equivalent and degenerate variants of the product, producing a heterogeneous microstructure known as {\em twinned martensite} \,\citep{kaubk}, containing alternate arrays of elastically coupled variants or twins. At the martensite - austenite interface, crystallographic incompatibility is accommodated {\em on the average}. This is one of the various possibilities of accommodating incompatibility and there is an extensive literature on experiments \,\citep{cahn,kaubk,olson}, microscopic and coarse grained theory \,\citep{krum1,jac,krum2,drop,strn-only1,prl2,strn-only2,paper2,porta} and coarse-grained and atomistic simulations \,\citep{physica,prl2,paper1,kastner1,kastner2,space} of such martensites. 

On the other hand, there exists many other kinds of martensitic microstructures that incorporate defects, such as dislocation arrays \,\citep{mardis}, in order to make the interface compatible on the average. The incompatibility problem may also be resolved by large scale diffusion of particles, as is the case in 
ferritic microstructures which involve no twins and are produced at higher temperatures where solid-state diffusion is appreciable\, \,\citep{cahn,mullinsfer}. 
In a realistic situation, multiple modes of accommodating interfacial strain compatibility may be present. Though there have been attempts to include the effect of isolated defects within an otherwise strain-only approach\, \,\citep{groger}, a complete theory of microstructure, capable of describing dislocated, twinned, as well as ferritic microstructures has not yet been attempted.

It is clear that one needs to incorporate additional degrees of freedom, other than strains, in order to describe multiple microstructures within a single framework. Accordingly, we include contributions from non-elastic, non-affine excitations\, \,\citep{kroner,egami}; i.e., local rearrangements of atoms which cannot be represented as a coarse-grained strain (Fig.\,\ref{phi}(a)).
Detailed molecular dynamics simulations of early time nucleation in a solid undergoing a (square $\to$ rhombic) structural transition\, \,\citep{paper1} show that $(1)$ non-affine strains are generated during the transition and are localized in regions we call {\em non-affine zones} (NAZ), $(2)$ the transformation is driven by a small number of {\em active} particles\, \,\citep{space} undergoing rearrangements within NAZs, $(3)$ the NAZs lie close to regions where the non-order parameter (NOP) strains are larger than a threshold and lastly, $(4)$ the dynamics of NAZ determine microstructure selection so that a martensite forms when the NAZs are localized at the interface while a ferritic microstructure forms when the NAZs spread over large scales. 


We had earlier described microstructure selection in solid state transformations, using a Landau theory which goes beyond strain-only approaches by including both elastic and non-affine degrees of freedom $\phi$ that explicitly violate compatibility  \,\citep{paper2}. 
We had shown that our theory could describe both martensitic and ferritic microstructures, and derive conditions when one or the other obtain. In this paper, we  describe a droplet theory\, \,\citep{drop,porta} to study the structure and growth of a small region of martensite embedded in the matrix of the parent austenite phase. Simple but realistic variational ans{\"a}tze for the order parameter and non-affine fields enables us to derive the optimized free energy of this droplet and show that there are multiple solutions corresponding to twinned and dislocated martensites within specific ranges of system-dependent parameters.    

The plan of the paper is as follows. In section\,\ref{formulation}, we discuss the dynamical equations for the elastic and non-affine variables. In section\,\ref{lyapunov}, we show that the equations of motion admit a Lyapunov functional, which we use to explore the space of solutions. In section\,\ref{droplet} we describe the droplet variational ansatz of martensite embedded in the austenite phase. In section\,\ref{solutions} we show that, in the space of our system-dependent external parameters the Lyapunov functional has multiple minima which may be related to structurally and dynamically distinct microstructures. We discuss our results in section\,\ref{discussion} in the light of known results and conclude.



\section{Dynamical equations for elastic and non-affine variables}
\label{formulation}
In this section, we will paraphrase our earlier derivation \, \,\citep{paper2} of the dynamical equations for the elastic strains and $\phi$ during a structural transition. While our theory is general, motivated by our earlier work\, \,\citep{paper1,paper2,space}, we describe a formulation based on a square to rhombic transformation. The square to rhombic transition is a special case of the more general transition from square to oblique two dimensional Bravais lattices. The affine shear strain,  $e_3 = \partial u_x/\partial y + \partial u_y/\partial x$ and the affine deviatoric strain $e_2 = \partial u_x/\partial x - \partial u_y/\partial y$ are the order parameters (OP) for the square to oblique transition, where $u_i (i=x,y)$ is the displacement field describing deformations from the (square) reference lattice. The equilibrium value of $e_2$ for the rhombic lattice is zero. The volumetric affine strain $e_1^A = \partial u_x/\partial x + \partial u_y/\partial y$  is a non-order parameter (NOP) strain which relaxes much faster than the OP strains\, \,\citep{paper2} and will be taken to be slaved to the latter. 

\begin{figure}[t]
\begin{center}
\includegraphics[width=14.cm]{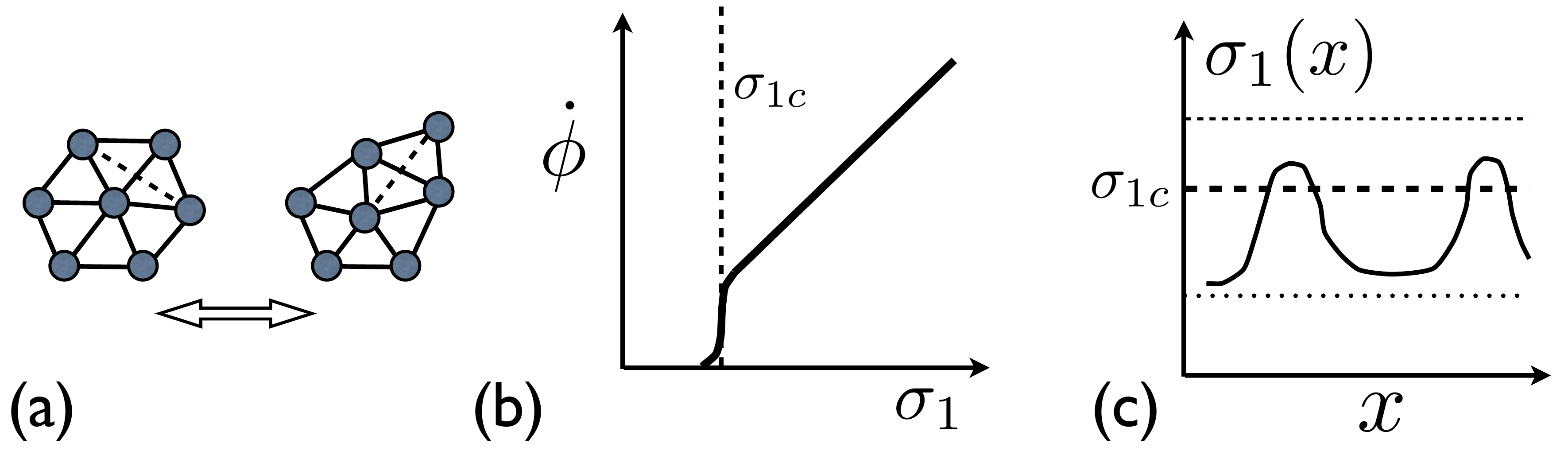}
\end{center}
\caption{(a) Schematic illustration of a typical local rearrangement producing non-affine displacements. Note that such a transformation produces $5$ and $7$ coordinated neighbourhoods which may be interpreted as a pair of tightly packed dislocations with overlapping cores. (b) Illustration of the threshold dynamics for the non-affine field given in (\ref{micplas2}). (c) Schematic illustration of NAZ produced when the stress $\sigma_1(x)$ crosses the threshold $\sigma_{1c}$. Horizontal lines show three values for $\sigma_{1c}$. For large $\sigma_{1c} > \sigma_1(x)$ (thin dashed line) everywhere, NAZ are never produced. As $\sigma_{1c}$ decreases, small isolated NAZs begin to form --- the regime described in our present calculation (bold dashed line). For $\sigma_{1c}$ which is too small (dotted line), NAZs form everywhere and our droplet ansatz eventually breaks down.}
\label{phi}
\end{figure}

We incorporate non-affine fields in our description by assuming that the {\em total} NOP strain may be decomposed into an affine $e_1^A$ and a non-affine part $\phi$, i.e., $e_1 = e_1^A + \phi$. This decomposition is analogous (though not equivalent !) to the decomposition of total strain into elastic and plastic parts, routinely made in the continuum theories of plasticity\, \,\citep{lubarda,acharya}. In contrast to most plasticity theories, $\phi$, is however generated in the {\em volumetric} sector, an  unusual requirement because deformation in solids is known to remain elastic for large compressional or dilatational stresses with plasticity being produced mainly under shear deformation, $e_3$. This requirement can be readily understood, if we remind ourselves of the following : (1) the shear strain $e_3$, which is the broken symmetry OP strain, is large within the martensite droplet and decays to zero in the parent phase, (2) gradients of $e_3$ generate large NOP ($e_1^A$) strain at the interface in order to satisfy elastic compatibility and (3) the combined presence of $e_3$ and $e_1$ makes the interfacial region susceptible to local rearrangements of particles, this generates NAZs \, \,\citep{paper1} in regions of high NOP strain present at the moving parent-product front. The topological rearrangements, now parametrized by $\phi$, occur when the local conjugate stress $\sigma_1(\{\epsilon_3\})$ exceeds an appropriate local threshold $\sigma_{1c}(\{\epsilon_3\})$ or yield criterion. Note that both the conjugate stress and the threshold which determine $\phi$ are finally {\em functionals} of $e_3$.

The Lagrangian is given by,
\begin{eqnarray}
{\cal L} &=& \int \frac{\rho}{2}\left[(\dot u_x^2 + \dot u_y^2)\right]dxdy - {\cal F},
\label{lag0}
\end{eqnarray}
where the first term is the kinetic energy with $\rho$ the mass density, which may be set to unity without loss of generality.
The free energy functional ${\cal F}$ depends on the structural transition being considered and for the two-dimensional square to rhombic transition, one may set,
\begin{eqnarray}
\label{freen1}
{\cal F}  &=&  \frac{1}{2}\int\,{\rm dxdy}\,\left[ a_1(e_1^A+\phi)^2\,+\,a_2e_2^2\,
+ a_3e_3^2 -\, \frac{1}{2}b_3e_3^4 
     \right. \nonumber \\  
&+& \frac{1}{3}d_3e_3^6 + c_1(\nabla e_1^A)^2\, +\,\left.c_2(\nabla e_2)^2\,+\,c_3(\nabla e_3)^2  \right].
\end{eqnarray}
The parameters $a_1, a_2$ and $a_3$ are elastic constants appropriate for the square (parent) phase and $c_1, c_2$ and $c_3$ are related to strain correlation lengths. For the range of parameters considered, ${\cal F}$ has three minima at $e_3 = 0$ representing the square, parent or austenite and $e_3 = \pm e_0 = \pm (b_3+\sqrt{b_3^2-4 a_3 d_3}/2 d_3)^{1/2}$, the two degenerate variants of the product rhombic phase. The connection with external control parameters such as  temperature ($T$) is, as usual, through the temperature dependence of the coefficients.

 To obtain the equation of motion in the displacement fields, we need to solve the Euler Lagrange equation,
\begin{equation}
\frac{d}{dt}\frac{\partial {\cal L}}{\partial \dot u_i} 
- \frac{\partial {\cal L}}{\partial u_i} = 
- \frac{\partial {\cal R}}{\partial \dot u_i}.
\label{euler-lag}
\end{equation}
The Rayleigh dissipation functional  \,\citep{landau} ${\cal R} = \frac{1}{2}\int \left[\gamma_1 (\dot e_1^A)^2 + \gamma_2 \dot e_2^2 + 
\gamma_3\dot e_3^2\right]dxdy$, where the coefficients $\gamma_1, \gamma_2$ and $\gamma_3$ are the 
corresponding ``viscosities''.

We take $e_1^A$ to be a slaved variable which reaches steady state much faster than the OP strains. The equations of motion for the components of the
strain can then be written as,
\begin{eqnarray}
\label{e1-dyn}
\nabla^2 e_1^A &=& q_{13}\frac{\partial^2e_3}{\partial x\partial y} 
- \nabla^2 \phi, \\
\label{e3-dyn}  
\ddot e_3 &=& \nabla^2\Big( a_3e_3 - b_3e_3^3 + d_3e_3^5  
- c_3\nabla^2e_3 - 2c_1\frac{\partial^2 e_1^A}{\partial x\partial y} + \gamma_3\dot e_3 \Big) \nonumber \\
& + & 2\frac{\partial^2}{\partial x\partial y}\Big\{a_1(e_1^A+\phi)\Big\},
\end{eqnarray}
where $q_{13}=2(a_2-a_3)/(a_1+a_2)$. The OP strain $e_2$ has been eliminated using a modified St. Venant condition,
\begin{eqnarray}
\nabla^2 e_1^A - \left(\frac{\partial^2}{\partial x^2} - \frac{\partial^2}{\partial y^2}\right)e_2 - 2\frac{\partial^2e_3}{\partial x\partial y} = -\nabla^2 \phi.
\label{venant}
\end{eqnarray}
where the strain incompatibility is completely determined by gradients of $\phi$. To obtain a closed set of equations, we need to prescribe a dynamics
for the non-affine field $\phi$. To arrive at this equation,
we note that $\phi$ incorporates two dynamical processes. First, $\phi$ is produced at a rate given by $h_1^{-1}$, whenever $\sigma_1$ locally exceeds a threshold $\sigma_{1c}$. We denote the nonlocal response kernel as $K(t,t')$, where $t,t'$ are times. This is the source of NAZs (Fig.\ref{phi}(b)), and goes to reduce the local stress $\sigma_1$. Second, as the transformation proceeds, $\phi$ produced at regions of large $\sigma_1$ (Fig.\ref{phi}(c)) at the austenite-martensite interface, needs to anneal diffusively with coefficients  $c_x'$ and $c_y'$. The martensite droplet, being a broken symmetry phase automatically defines an interface or habit plane along which elastic compatibility is restored on the average. In general, atomic mobilities are different in the directions parallel and perpendicular to this interface suggesting $c_x' \neq c_y'$. 
These processes suggest the following phenomenological equation of motion for $\phi$,
\begin{eqnarray} 
\label{micplas2}
\dot \phi&=&-\int_{-\infty}^t\frac{\sigma_1}{h_1}H\left(|\sigma_1|-\sigma_{1c}\right)K(t,t')dt^\prime\nonumber \\
&+& c_x'\Big(\frac{\partial^2\phi}{\partial x^2}\Big) +  c_y'\Big(\frac{\partial^2\phi}{\partial y^2}\Big).
\end{eqnarray}
In the above equation, the local volumetric stress $\sigma_1 = \delta  {\cal F}/\delta e_1^A = a_1(e_1^A+\phi)$, $H(x)$ is a unit step function and we have chosen a simple threshold criterion with a constant yield stress $\sigma_{1c}$, independent of any state variable (Fig.\ref{phi}(b)). The  phenomenological parameters $h_1, c_x'$ and $c_y'$ in (\ref{micplas2}), may, in principle, be obtained from careful computer simulations of detailed  atomistic models.  

\section{The Lyapunov functional}
\label{lyapunov}
The equations (\ref{e1-dyn}) and (\ref{e3-dyn}) together with (\ref{micplas2}) describe the dynamics of $\epsilon_i$ ($i=1,3$) and $\phi$ completely. We are interested in the steady state solutions of these equations. While the equations of motion of the strains is derived from the free energy functional (\ref{freen1}), the non-affine field follows a phenomenological dynamics (\ref{micplas2}) which does {\em not} arise from this free energy. In this section we show that, nevertheless, one may be able to define a Lyapunov functional whose minimizers do correspond to steady states of (\ref{e1-dyn}), (\ref{e3-dyn}) and (\ref{micplas2}) thereby reducing our dynamical problem to one of optimization.

Using $K(t,t') = \delta(t-t')$ to simplify the numerics, the phenomenological dynamics of the $\phi$ field becomes,
\begin{eqnarray}
\dot \phi&=& \frac{\sigma_1}{h_1}H\left(|\sigma_1|-\sigma_{1c}\right) 
+ c_x'\Big(\frac{\partial^2\phi}{\partial x^2}\Big) +  c_y'\Big(\frac{\partial^2\phi}{\partial y^2}\Big), 
\label{phidyn}
\end{eqnarray}
where $H(\sigma)$ is the Heaviside function. Equation (\ref{phidyn}) needs to be solved under the boundary conditions $\phi(t=0) = 0$ and $\phi \to 0$ at large distances and reduces to two steady state situations depending on the magnitude of $\sigma_1$. For $|\sigma_1| > \sigma_{1c}$ we have, 
\begin{equation}
- \frac{a_1 (e_1^A + \phi)}{h_1} =   c_x'\Big(\frac{\partial^2\phi}{\partial x^2}\Big) +  c_y'\Big(\frac{\partial^2\phi}{\partial y^2}\Big)
\end{equation}
while for $|\sigma_1| < \sigma_{1c}$ we obtain the anisotropic diffusion equation,
\begin{equation}
c_x'\Big(\frac{\partial^2\phi}{\partial x^2}\Big) +  c_y'\Big(\frac{\partial^2\phi}{\partial y^2}\Big) = 0.
\end{equation}
Both these solutions may be combined as a single equation,
\begin{equation}
c_x\Big(\frac{\partial^2\phi}{\partial x^2}\Big) +  c_y\Big(\frac{\partial^2\phi}{\partial y^2}\Big) + a_1 (e_1^A + \phi)f(\Lambda,\sigma_{1c},\sigma_1) = 0
\label{arb}
\end{equation}
where $f(\Lambda,\sigma_{1c},\sigma_1)$ belongs to a one-parameter family of smooth functions which represents the Heaviside function in the limit $\Lambda \to 0$. For our computational convenience, we choose  
\begin{equation}
c_x\Big(\frac{\partial^2\phi}{\partial x^2}\Big) +  c_y\Big(\frac{\partial^2\phi}{\partial y^2}\Big) + a_1 (e_1^A + \phi) \Big\{1-\exp\Big(-\frac{|\sigma_1|}{\sigma_{1c}}\Big)\Big\} = 0
\label{reps}
\end{equation}
where we have absorbed the constant $h_1$ into the definition of the dynamical parameters $c_x$ and $c_y$ quantifying the ratios of the times it takes for $\phi$ to form to that it takes to spread in the $x$ and $y$ directions respectively. One can show by direct plotting that the chosen representation for the step function, though strictly accurate for small $\sigma_{1c}$, reproduces expected qualitative behaviour even if $\sigma_{1c}$ is larger. We emphasize that this choice is not a 
small $\sigma_{1c}$ approximation, in principle we could have chosen a different smooth representation for the step function, accurate over a larger
range of $\sigma_{1c}$.

 We now seek a Lyapunov functional whose minimizers would automatically generate the steady state solutions for $\epsilon_i$ and $\phi$. It is straightforward to show that this corresponds to,  
\begin{eqnarray}
{\cal F}_L  & = & \frac{1}{2}\int \Big\{ a_1(\phi^2 + 2 e_1^A\phi) + 2 \Big(\frac{\sigma_{1c}}{a_1}\Big)( \sigma_{1c}+\sigma_1) e^{-\frac{|\sigma_1|}{\sigma_{1c}}}\nonumber \\
& & + c_x\left(\frac{\partial \phi}{\partial x}\right)^2 + c_y\left(\frac{\partial \phi}{\partial y}\right)^2 \Big\} dxdy \nonumber \\
& & + \dots
\end{eqnarray}
where the dots represent terms which are independent of $\phi$. Collecting terms we obtain,
\begin{equation}
{\cal F}_L  = {\cal F} + \frac{1}{2}\int \Big\{\frac{2 \sigma_{1c}}{a_1}( \sigma_{1c}+\sigma_1)\,e^{-\frac{|\sigma_1|}{\sigma_{1c}}} 
+ c_x\left(\frac{\partial \phi}{\partial x}\right)^2 + c_y\left(\frac{\partial \phi}{\partial y}\right)^2 \Big\} dxdy.
\label{lyapu}
\end{equation}

One can check that ${\cal F}_L$ is, indeed, locally positive definite, and the system relaxes to the local minima of ${\cal F}_L$; the basic requirements of a Lyapunov functional\, \,\citep{mazur,lyapun}. Note however that the Lyapunov functional is not unique. Indeed, in a closed system the irreversible entropy production rate may itself serve as an appropriate Lyapunov functional\,\,\citep{nieto}.

We can now readily verify that a functional minimization of ${\cal F}_L$  obtains the required steady state solutions for strains and $\phi$. Finally, note that in (\ref{lyapu}), the first term within the integral is a purely elastic contribution whereas the second term is nonaffine. Elasticity favors the production of $\phi$ (with appropriate sign) once the threshold criterion is satisfied, since the elastic energy is reduced through stress relaxation. On the other hand, large $c_x$ and/or $c_y$ inhibits the formation of (nonuniform) $\phi$ at the NAZ  where stress is large. The balance of these two tendencies optimize the magnitude and the sign of $\phi$ and, consequently, the nature of the microstructure. 
\begin{figure}[t]
\begin{center}
\includegraphics[width=11cm]{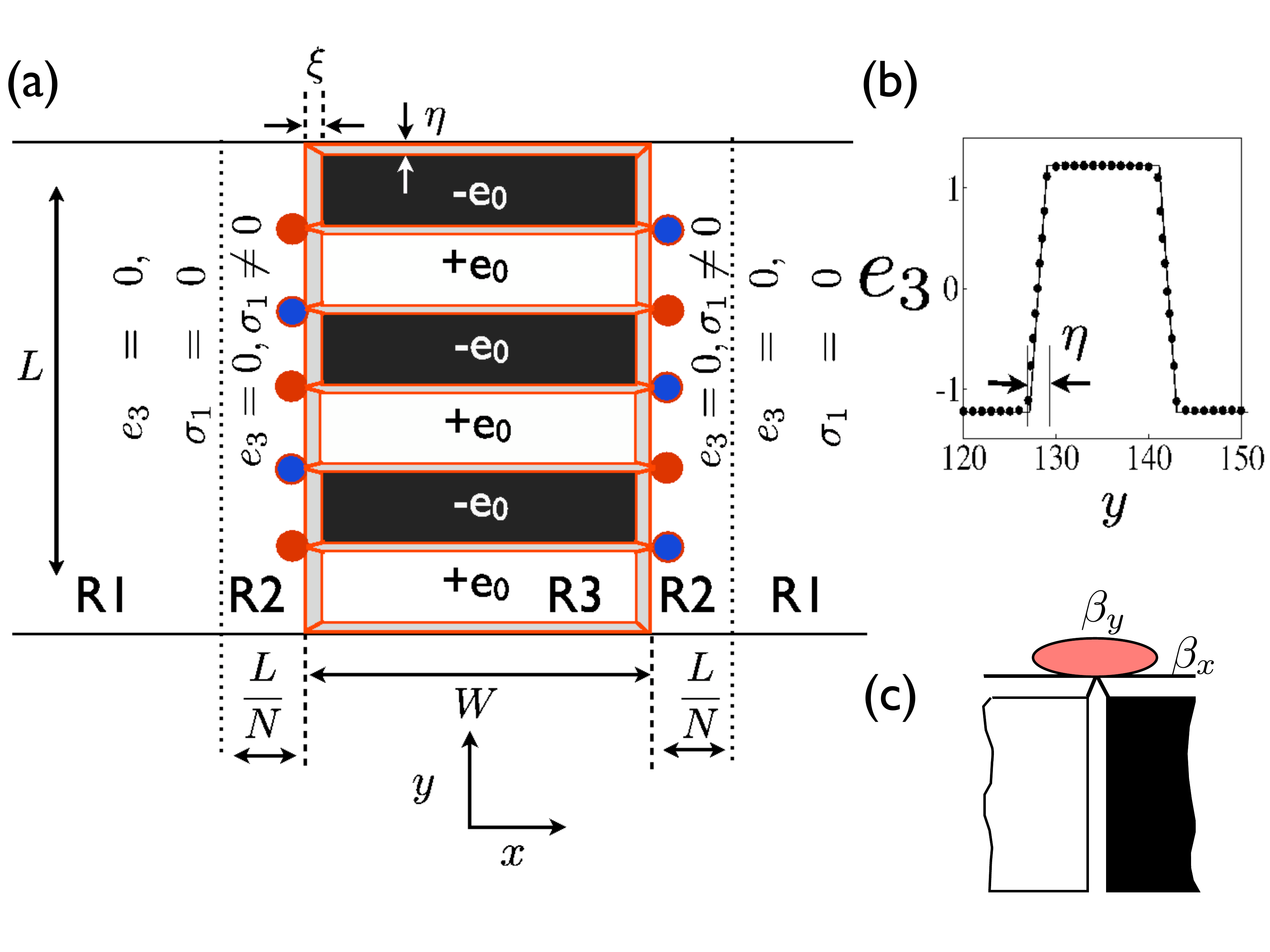}
\end{center}
\caption{(a) Schematic diagram showing our ansatz for $e_3$ corresponding to the martensite droplet. Periodic boundary conditions are assumed in both the $x$ and $y$  direction.There are three regions $R1,R2$ and $R3$. The region $R1$ contains the parent austenite ($e_3=0$). All strain and associated stress fields are zero in this region. The region $R3$ is of size $L \times W$, and contains ${\cal N}$ twins (6 shown). The twins consist of regions where $e_3 = \pm e_0$, designated by white and black, which are degenerate in free energy. The gray regions are interfaces with widths $\eta$ and $\xi$. The region $R2$ has no $e_3$ but contains a non zero non-order parameter strain $e_1^A$ (and associated stress $\sigma_1$) which is produced due to the strain inhomogeneity created at the droplet. $\sigma_1$, which alternates in sign, reaches maximum absolute values at the blue and red colored circles marking regions of maximum strain inhomogeneity. (b) The ansatz $e_3(y)$ vs $y$  (solid line) and the numerical solution of the dynamical equation (filled squares). The interfacial width $\eta$ is shown. (c) A close up of the triple junction where the two twin variants (white and black) meet at a austenite-martensite interface. The red ellipse with widths $\beta_x$ and $\beta_y$ represents a typical NAZ.}
\label{ansatze}
\end{figure}

\section{The Droplet Ansatz}
\label{droplet}

Our variational ansatz for $e_3(x,y)$ is shown in Fig.\ref{ansatze}(a). We consider a rectangular droplet of the martensitic (twinned rhombic) phase occupying a region $R3$ of dimension $L\times W$ and consisting of a linear array of ${\cal N}$ twins with alternating values of the order parameter $e_3 = \pm e_0$ represented by alternating strips of black and white. The droplet is oriented such that the {\em habit planes} or the parent product interfaces coincide with the lines $x = \pm W/2$ and the twin interfaces $y_k = -L/2 + k L{\cal N}$, for $k = 0 \ldots {\cal N}$. We assume periodic boundary conditions in both directions. All interfaces are assumed to be flat which we later show to be a reasonable assumption for most cases. The region $R1$ represents the parent austenite (square) phase $e_3 = 0$ which is unaffected by the presence of the martensite droplet. Within the region $R2$ of size $\sim L/{\cal N}$, the order-parameter $e_3$ still vanishes but the presence of the droplet produces stress fields $\sigma_1$ which determines the structure and dynamics of the interface. In Fig.\ref{ansatze}(b) we have shown a typical interface of width $\eta$ between twins with $e_3 = +e_0$ and $e_3 = -e_0$.  The points are from a numerical solution of the dynamical equations (\ref{e1-dyn})\,\citep{paper2}, and the line is a linear interpolation which we use for our droplet ansatz here. The interface between the parent and the product where the order parameter $e_3$ interpolates between $\pm e_0$ and $0$ is similarly represented by a linear ansatz with interfacial width $\xi$. Within the region $R2$, the order parameter $e_3$ vanishes and hence non-linearities in the free energy functional may be neglected. 

Since the volumetric strain is not a broken symmetry variable, it relaxes rapidly to its steady state value determined by the configuration of $e_3$. We  determine
$e_1^A$ by solving the partial differential equation (\ref{e1-dyn}) in the region $R2$. Equation (\ref{e1-dyn}) is essentially a Poisson equation with ``electric potential''  $e_1^A$ being produced by a ``charge density'' $\partial^2e_3/\partial x\partial y$ with the boundary condition $e_1^A = 0$ at distances large from the droplet, i.e., region $R1$. In region $R2$, $\partial^2e_3/\partial x\partial y = 0$ everywhere except at the triple junctions where twin interfaces meet the martensite-austenite interface (shown in Fig.\ref{ansatze}(a)\&(c)). This analogy is expected to be valid in the presence of small or moderate amounts of $\phi$ which reduces the effect of the anisotropic long range elastic kernel. This  corresponds to a situation when the threshold stress is relatively large together with small diffusion times so that $\phi$ accumulated at the interface is never too large (Fig.\ref{phi}(c)). When, on the other hand, generous amounts of $\phi$ are produced, this relation no longer holds. Therefore, in the limit of small interfacial width, $e_1^A$ is the electric potential in 2D due to a configuration of {\em point} charges of alternate sign $\pm \sqrt{2 \pi}\lambda$ centered at the triple junctions with, 
\begin{equation}
\label{lamb}
\lambda = \frac{1}{\sqrt{2 \pi}}\,q_{13}e_0\left(\frac{1}{\xi}+\frac{1}{\eta}\right)\sqrt{\xi^2+\eta^2}.
\end{equation}
For large ${\cal N}$, the potential ($e_1^A$), at any point $(x,y)$ within $R2\,\&\,R3$, due to this charge configuration in absence of $\phi$ may be written as 
\begin{eqnarray}
\label{affine}
e_1^A(x,y) =
\lambda\cos\left(q\,y\right)\left(e^{-q|x+W/2|}-
e^{-q |x-W/2|}\right) 
\end{eqnarray}
where $q=\pi {\cal N}/L$ and we have used a Fourier decomposition limiting ourselves to the most significant term. 

Initially, the stress $\sigma_1 = a_1 e_1^A$ is also largest at these triple junctions and is most likely to cross the threshold $\sigma_{1c}$ there generating $\phi$ (which decreases $\sigma_1$). Accordingly, we supplant our ansatz for the OP field with one for $\phi$ viz.  
\begin{equation}
\phi = \phi_0\left[e^{-(x+W/2)^2/2\beta_x^2} - e^{-(x-W/2)^2/2\beta_x^2}\right]\sum_{k=-{\cal N}}^{\cal N}e^{-(y+kL/2{\cal N})^2/2\beta_y^2}\cos(qy)
\end{equation}
which peaks at every triple junction with anisotropic widths $\beta_x$ and $\beta_y$ in the $x$ and $y$ directions respectively and an amplitude which changes sign following that of $e_1^A$. The form of the ansatz (shown schematically in Fig.\ref{ansatze}(c)) interpolates between a $\phi$ which may either be strongly localized (small $\beta_x, \beta_y$) or one which completely wets the austenite~-martensite interface and is modulated with a period matching that of $e_1^A$ and $e_3$.

Substituting the OP and slaved NOP strains and $\phi$ in ${\cal F}_L$ we obtain the droplet free energy 
\begin{eqnarray}
{\cal F}_L = I_bLW + 2\Sigma_{pp}L + \Sigma_{tw}W {\cal N}+ 2\pi a_1\lambda^2 \frac{L^2}{\cal N} + F_{\phi}(\phi_0,\beta_x,\beta_y)\,{\cal N}
\label{drop-free}
\end{eqnarray}
where the constant  ${\cal I}_b = a_3e_0^2/2 - b_3e_0^4/4 + d_3e_0^6/6$ multiplies the first term representing the contribution from the bulk of the droplet. The second and the third terms with coefficients $\Sigma_{pp} = c_3\int_{-\infty}^{\infty}dx\left(\partial e_3/\partial x\right)^2 = 2e_0\sqrt{c_3{\cal I}_s}$, the surface tension due to the parent-product and $\Sigma_{tw} = \frac{c_3}{{\cal N}}\int_{-\infty}^{\infty}dy\left(\partial e_3/\partial y\right)^2 = 4e_0\sqrt{c_3{\cal I}_s}$ for each twin interface with ${\cal I}_s = a_3e_0^3/3 - b_3e_3^5/5 + d_3e_0^7/7$, represent the energies of the parent-product and twin interfaces.  The fourth term gives the contribution of the fringing NOP field $e_1^A$. The function $F_{\phi}(\phi_0, \beta_x,\beta_y)$ in the fifth term is the contribution of the non-affine field, $\phi$, to the Lyapunov functional ${\cal F}_L$ which we had to obtain by numerical integration, checking our results against analytical expressions obtained in the pathological but instructive limit $\sigma_{1c} \to 0$ (see Appendix). $F_{\phi}(\phi_0, \beta_x,\beta_y)$ contains terms which are linear and quadratic in $\phi_0$. Unimportant contributions from gradients of $e_1^A$ have also been ignored for simplicity.

Equation (\ref{drop-free}) is  minimized with respect to $\beta_x,\,\beta_y,\,{\cal N}$ and $\phi_0$ as well as the interfacial widths $\xi$ and $\eta$ for a fixed choice
of all the parameters and size of the droplet $L$ and $W$. 
\begin{figure}[t]
\begin{center}
\includegraphics[width=14cm]{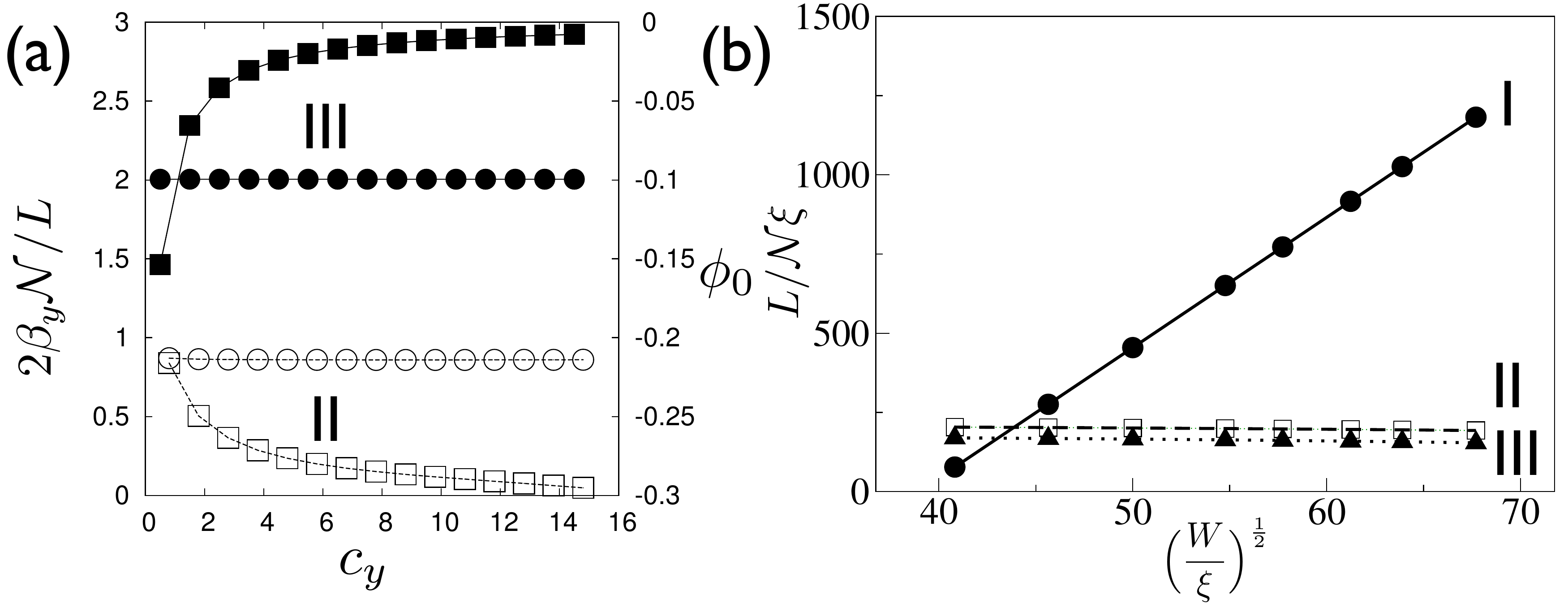}
\end{center}
\caption{(a) Plot of the scaled width $\beta_y$ of the $\phi$ distribution for {\hv  II} and {\hv  III} (open and filled circles, axis on the left) and $\phi_0$ (open and filled squares, axis on the right) as a function of $c_y$ for $c_x$ at the triple point shown in Fig.\ref{2dpd}(a) for $\sigma_{1c} = .001$. (b) Typical plots of $L/{\cal N}$ as a function of $W^{1/2}$ scaled by the interfacial width $\xi$ for a set of parameters over which each solution is stable. Bold line: solution {\hv I} with $a_1 = 0.05$, $\lambda = 0.1$, $c_x = 0.5$ and $c_y = 4$; dashed line: solution {\hv II} with $a_1 = 1$, $\lambda = 1.$, $c_x = 0.1$ and $c_y = 1$;
dotted line: solution {\hv III} with $a_1 = 1$ and $\lambda = 1.$, $c_x = 1$ and $c_y = 1$. Note that while solution {\hv I} reproduces the behavior of ferro-elastic martensites showing strong elastic coupling between the twins, in solution {\hv  II} and {\hv  III} the twins are elastically decoupled.}
\label{loc}
\end{figure}

\section{Steady state solutions and the dynamical phase diagram}
\label{solutions}


Minimization of ${\cal F}_L$ with respect to the interfacial widths immediately yields $\eta = \xi = e_0\sqrt{c_3/{\cal I}_s}$ giving $\lambda = 2q_{13}e_0/ \sqrt{\pi}$. Since we know that interfacial widths in solids are typically of the order of a lattice spacing\, \,\citep{kaubk}, this correspondence sets the microscopic length scale of our continuum model. Using these values, we minimize  ${\cal F}_L$ with respect to the rest of the free parameters, viz., $\beta_x, \beta_y, \phi_0$ and ${\cal N}$ for a range of values for  $c_x, c_y$ and  $\sigma_{1c}$. Our main results may be summarized as follows.
\begin{enumerate}

\item  We obtain three distinct solutions or {\em dynamical phases}, labelled {\hv I, II} and {\hv  III} which correspond to unique local minima of ${\cal F}_L$  in particular regions of parameter space.

\item The three dynamical phases show a wide variation in physical properties. While solution {\hv I} is essentially affine, solutions {\hv II} and {\hv III} involve non zero values of $\phi$.

\item   $\phi$, if present, is always strongly localized perpendicular to the habit plane, i.e. along $x$ so that $\beta_x  < \beta_y \ll W$ for all three solutions. Twin interfaces, therefore, do not ever contain any $\phi$. 

\item In solution {\hv II} $\phi$ is localized at regions of high stress concentration at the habit plane, while in {\hv III}, $\phi$ completely ``wets'' the habit plane.

\item Individual twins in {\hv I} are elastically coupled while they are decoupled in {\hv II} and {\hv III}. 

\item While {\hv I} is identified as twinned, ferro-elastic martensite, we propose that {\hv II} and {\hv III} are dislocated martensites.

\end{enumerate}

We now explain each of these results in detail. The magnitudes of $\phi_0$ and $\beta_y$ are controlled by the dynamical constants $c_y$ and $c_x$ and $\sigma_{1c}$.  For fixed $\sigma_{1c}$, solution {\hv I}, which exists for large $c_x$ and $c_y$, has negligible non-affineness ($\phi_0 \approx 0$) except at a small region of size $\sim \xi$ at the triple junctions. In this case, a suitable coarse graining of the interface taking $\xi \to 0$ may be used to derive a re-normalized strain-only theory without $\phi$  \,\citep{paper2}. Therefore, this solution most resembles ferro-elastic martensite domains describable completely within strain-only approaches \,\citep{strn-only1,strn-only2,porta}. 

To compare the nature of the two remaining solutions we have plotted $\beta_y$ and $\phi_0$ for both {\hv II} and {\hv III} as a function of $c_y$ for fixed $c_x$ over a typical range of parameters in Fig.\ref{loc}(a). The sign of $\phi_0$ in  both solutions {\hv  II} and {\hv  III} is {\em negative} showing that the non-affine field effectively decreases the NOP stress ($\propto \lambda $). Solution {\hv  II} comprises localized NAZ of large $|\phi_0|$ at the triple junctions while {\hv  III} has smaller $|\phi_0|$ which completely wets the habit plane. As $c_y$ increases, $|\phi_0| \to 0$ for solution {\hv III} though before it vanishes altogether, a first-order dynamical transition to {\hv I} intervenes. The width of the $\phi$ distribution along the habit plane, $\beta_y$ is relatively insensitive over the entire range of $c_y$ over which these solutions are stable.  

Increasing ${\cal N}$ leads to smaller NOP fringing fields $e_1^A$ at the austenite~-martensite interface and consequently lowers the contribution of the elastic energy due to elastic coupling between the twins. However, large ${\cal N}$ introduces more twin interfaces and hence increases the interfacial energy of droplet. For vanishing $\phi$, the interplay between these two factors alone decides the optimum number of twins ${\cal N}$. For example, we obtain
\begin{eqnarray}
\frac{L}{{\cal N}}&=& \sqrt{\frac{2 \pi \Sigma_{tw}W}{a_1 \lambda^2}}.
\label{scal}
\end{eqnarray}
This relation, between the size of the twin $L/{\cal N}$ and the width $W$, has been verified experimentally and has been obtained in earlier calculations  \,\citep{krum2,drop,porta}. Substituting the minimized ${\cal N}$ in (\ref{drop-free}), one obtains ${\cal F}_L$ as a sum of bulk ($\propto LW$) and surface ($\propto L+W$) terms with an effective surface tension given by $\Sigma_{eff} = \Sigma_{pp} + \sqrt{a_1 \lambda^2 \Sigma_{tw} W/2 \pi}$. 

Including $\phi$ changes the relation between the twin size and the width depending on the resulting optimum value for $\phi_0$ since $\phi$ changes both the numerator and the denominator of (\ref{scal}), viz.
 \begin{eqnarray}
\frac{L}{{\cal N}}&=& \sqrt{\frac{2 \pi (\Sigma_{tw}W + A \phi_0 + B \phi_0^2)}{a_1 \lambda^2 + 2 a_1 \lambda \phi_0}}.
\label{scalp}
\end{eqnarray}

Where $A$, and $B$ are positive definite functions of $\beta_x$, $\beta_y$ and $L/{\cal N}$. For $\phi \to 0$ viz. for solution ${\hv I}$, we can neglect the term quadratic in $\phi_0$, the small linear contributions to both the numerator and the denominator in (\ref{scalp}) then keeps the scaling relation unaltered. This is shown in Fig\ref{loc}(b) where we have plotted $L/{\cal N}$ vs. $W^{1/2}$ in scaled coordinates by changing $W$ in our ansatz for a fixed $L$ and for sets of parameters where each of the phases are stable. As $|\phi_0|$ increases, the solution for $L/{\cal N}$ is not straightforward because the numerator is itself a (complicated) function of $L/{\cal N}$ and one needs to solve (\ref{scalp}) self consistently and numerically. The result, also shown in Fig.\ref{loc}(b), reveals that $\phi$ decreases stress, since non-affineness {\em screens} the elastic interactions between twins. For both solutions {\hv  II} and {\hv  III}, the twins appear to be completely decoupled elastically suggesting that $\phi_0 \sim W$ canceling the $W$ dependence of the numerator making the twin width $L/{\cal N}$ almost independent of $W$. 
\begin{figure}[t]
\begin{center}
\includegraphics[width=14cm]{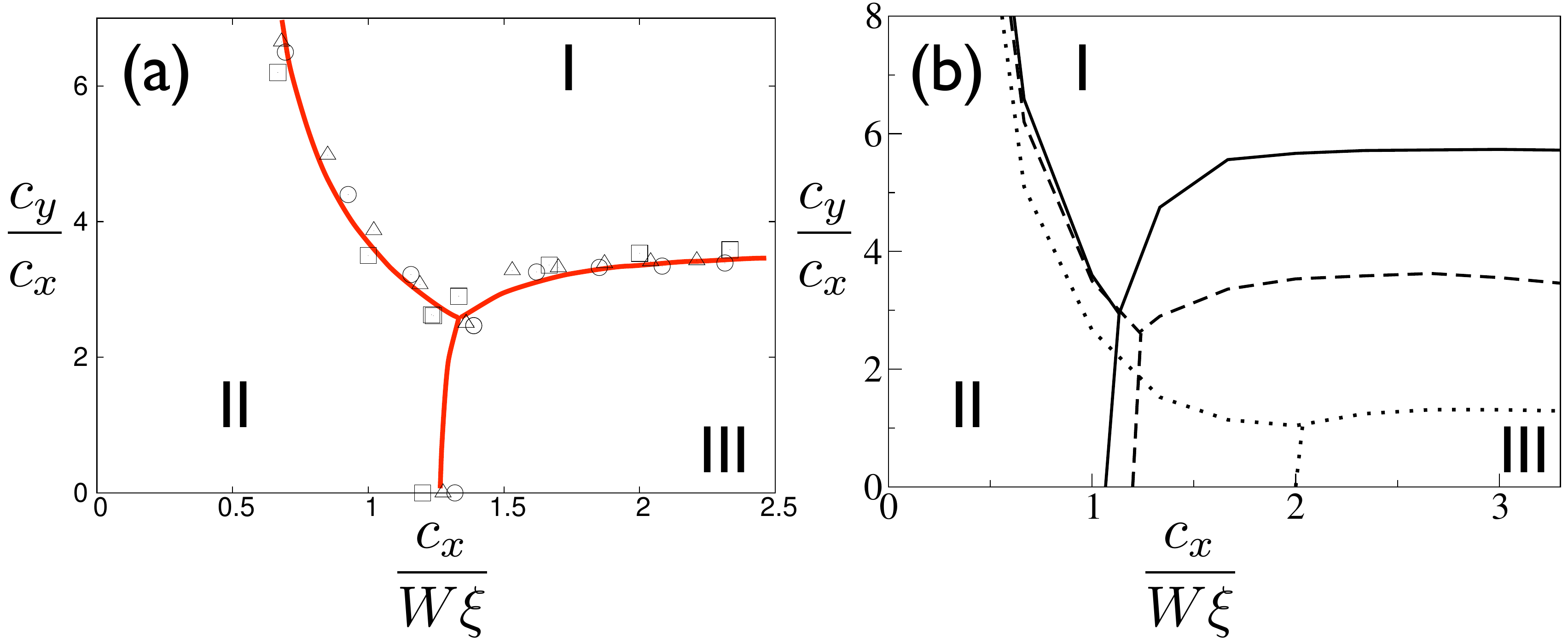}
\end{center}
\caption{(a) Dynamical phase diagram in the space of the scaled parameters $c_x/W\xi$ (see text) and $c_y/c_x$ showing relative stabilities of solutions {\hv I,II} and {\hv III} (see text) for three values of the twin width $W/\xi = 2500$ ($\Box$), $3600$ ($\circ$) and $4900$ ($\Delta$). Note that in the scaled coordinates, the phase boundaries (shown as red/gray lines) for different $W$ collapse on a single set of three curves. The threshold stress $\sigma_{1c} = .001$ (b) Dynamical phase diagram for three values of threshold $\sigma_{1c} = 0, .001$ and $.0025$ for $W/\xi = 2500$. Increasing $\sigma_{1c}$ stabilizes solution {\hv I} depressing the triple point and eventually removing {\hv II} and {\hv III} altogether. For $\sigma_{1c} \gtrsim  0.005$ the triple point obtains for the unphysical $c_y < 0$. The rest of parameters common for both (a) and (b) are $L/\xi = 10^5$, with $ a1 = 1., a2 = 0.4, a3 = 1.25, b3 = -5, c3 = 10^{-5}$ and $d3 = 5.$.} 
\label{2dpd}
\end{figure}

We obtain a dynamical phase diagrams showing regions of stability (lowest ${\cal F}_L$) for solutions {\hv I}, {\hv II}, and {\hv III} in  Figs.\ref{2dpd}(a) and (b) 
in the scaled parameters $c_x/W\xi$ and $c_y/c_x$ for fixed $\sigma_{1c}$. Remarkably, in these variables, the dependence on $W$, the size of the droplet, may be scaled away. For large $c_x$ and $c_y$ the system finds it increasingly difficult to generate non-affinity and there is a transition from structures {\hv  II} or {\hv  III} (large $|\phi_0|$) to {\hv I} (small $|\phi_0|$). Solids which form only twinned martensitic microstructures are expected to have dynamical parameters corresponding to this region of our phase diagram. For lower $c_x$ and $c_y$, on the other hand, $\phi$ is produced and solutions {\hv II} and {\hv III} are stable. In this region, as $c_x$ is increased, there is a jump in the values of both the localization parameter $2 \beta_y {\cal N}/L$ and $\phi_0$ as the system switches from solution {\hv  II} with localized $\phi$ with large amplitude to solution {\hv  III} with a relatively delocalized, broad (in $y$) $\phi$ distribution albeit with a smaller amplitude. During this transition from {\hv II} to {\hv III}, $\beta_x$ also decreases, though this change is not as much as in $\beta_y$. 

Finally, if $\sigma_{1c}$ is increased, predictably, non-affineness decreases stabilizing the ``strain-only'' {\hv I} solution. In this case, the triple point in the phase diagram shifts to lower $c_y$ values and the phase boundary between the {\hv I} and {\hv II} phases shifts to the left (Fig.\ref{2dpd}(b)) and eventually tends to unphysical negative values for $c_y$ and $c_x$. For large $\sigma_{1c}$, again, one recovers the results of strain-only theories where twinned martensites with $\phi = 0$, i.e. solution {\hv I}, is the only stable dynamical phase.    
\begin{figure}[t]
\begin{center}
\includegraphics[width=7cm]{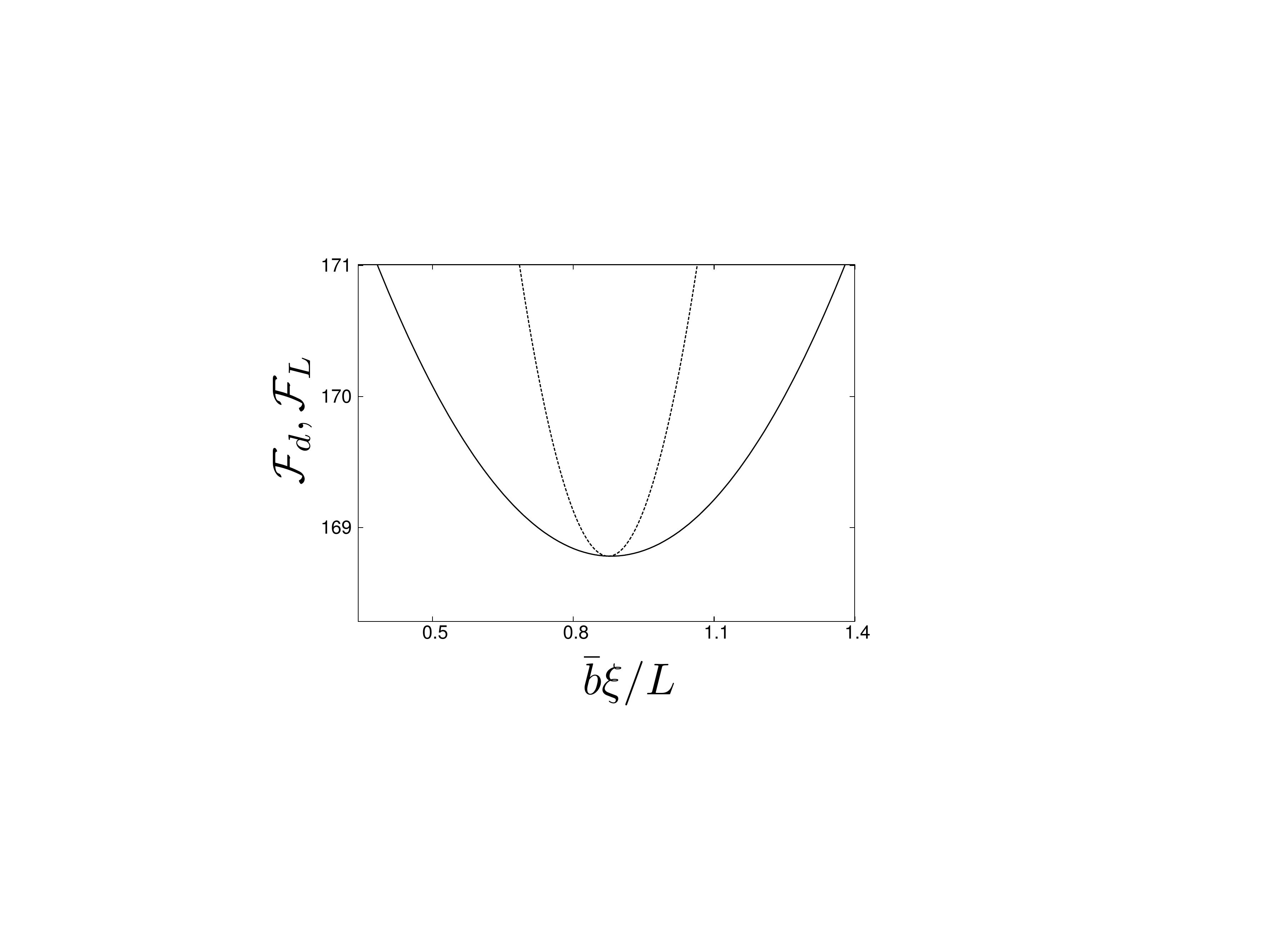}
\end{center}
\caption{Comparison of the free energy as a function of the scaled average Burgers vector density $\bar{b}\xi/L$ obtained from solution {\hv  II} (dotted line) and direct evaluation using an array of dislocations (bold line) for a droplet containing a single twin with $c_x$ and $c_y$ near the triple point values for $\sigma_{1c} = .001$. Note that the two curves yield comparable values for $\bar b$.}
\label{compare}
\end{figure}

While the interpretation of solution {\hv I} as a strain-only, ferroelastic, twinned martensitic droplet is clear, the nature of solutions {\hv  II} and {\hv  III} still needs to be established. Our calculation, shows that $\phi$ decreases the strong elastic coupling between the twins so that they are, in effect, individual grains of the product phase surrounded by an interface containing regions of large non-affinity where the St. Venant's incompatibility $\nabla^2 \phi \neq 0$. A non-affine region, defined in our work as a fluctuation in the local coordination number, may be interpreted as a compact droplet of dislocations with overlapping cores. Individual dislocations are easier to nucleate within such a region \, \,\citep{paper1} compared to the rest of the solid due to the lower barriers to dislocation nucleation within a NAZ. Normally the velocity of an austenite- martensite interface is large \, \,\citep{kaubk} and therefore nucleation of dislocations at the interface does not take place because nucleation times are too long. However, at high temperatures where the parent matrix is soft, and the interfacial velocity small, dislocation nucleation can and does take place with the strain mismatch across the parent product interface being accommodated by a lattice invariant strain composed of dislocations rather than twins. Such dislocated, ``massive'' martensites are commonly seen, for example, in low carbon, soft, iron alloys\, \,\citep{olson,mardis}. The movement of the product interface, or habit plane, is then accomplished by ``glide'' motion of the dislocations while movement of dislocations along the habit plane is due to the much slower ``climb'' motion.  

In order to investigate whether the presence of non-affine fields at the parent- product interface can be indeed related to interfacial dislocations, we use the correspondence suggested by  \,\citep{acharya} and  \,\citep{sethna} where the total strain is decomposed as $\overleftrightarrow{\mathbf e^{ }} =\overleftrightarrow{\mathbf e^A}+\overleftrightarrow{\mathbf e^p}$ i.e. consisting of a sum of affine, $\overleftrightarrow{\mathbf e^A}$, and plastic, $\overleftrightarrow{\mathbf e^p}$, parts. Accordingly we identify,  the incompatibility $\nabla\times\overleftrightarrow{\mathbf e^p}=\overleftrightarrow{\mathbf \zeta}$ with the dislocation density tensor which is given in component form,
$$
\zeta_{km} = \epsilon^{ijk}\frac{\partial e_{mj}^p}{\partial x_i},
$$
where $\epsilon^{ijk}$ is the unit antisymmetric tensor, the summation convention is implied and the indices run from $1-3$; for plane strain conditions relevant here, we need to look only at terms corresponding to $k=3$, viz. 
\begin{gather*}
\nabla\times \overleftrightarrow {e^p} =  \begin{pmatrix} \zeta_{31}\\ \zeta_{32}\end{pmatrix} = \frac{1}{2}\begin{pmatrix} -\frac{\partial e_1^p}{\partial y}\\ \frac{\partial e_1^p}{\partial x}\end{pmatrix} 
\end{gather*}
representing the $x$ and $y$ components of the Burgers vector density. To obtain the total dislocation density, we need to coarse grain $\zeta_{31}$ and $\zeta_{32}$ over a region $L\times\xi$. The $y$ component vanishes by symmetry yielding finally the mean Burgers vector content $\bar b$ per unit length of the interface. From the results of the last section, we conclude that $\bar b \approx 0$ for solution {\hv I}; the austenite- martensite interface in this case being virtually free of any dislocation content. This is not true for solutions {\hv  II} and {\hv  III} where we expect to obtain an optimum $\bar b$ related to the optimum profile for the $\phi$ field i.e. $\phi_0$, $\beta_x$ and $\beta_y$. Since all of these quantities are almost independent of $c_x$ and $c_y$ within the range over which each solution is stable, it, therefore, makes sense to compare $\bar b$ with that obtained from an array of  evenly spaced dislocations at the austenite- martensite interface each with Burgers vector ${\bar b}$ oriented along the $x$ direction \,\citep{mardis}. Considering that creating each dislocation needs an expenditure of core energy $E_{core} = a_3{\bar b}^2/4\pi$ one can obtain the Burgers vector density by optimizing the Landau free energy,    
\begin{eqnarray}
{\cal F}_d = \frac{1}{2}\int \left[a_1e_1^2 + a_2e_2^2 + a_3e_3^2 + \frac{1}{2}b_3e_3^4 + \frac{1}{3}d_3e_3^6 + c_3(\nabla e_3)^2\right]dxdy + N_dE_{core}
\end{eqnarray}
where the strains have contributions from the OP ($e_3(x,y)$) and NOP ($e_1^A(x,y)$) strains from the martensite droplet as well as from the array of interfacial dislocations with each dislocation (chosen to be at the origin) contributes,
 \begin{eqnarray}
\frac{\partial u_x}{\partial x} &=& -Dy\frac{3x^2-y^2}{(x^2+y^2)^2}\\
\frac{\partial u_y}{\partial y} &=& Dy\frac{x^2-y^2}{(x^2+y^2)^2}\\
\frac{\partial u_x}{\partial y} &=& \frac{\partial u_y}{\partial x} = Dx\frac{x^2-y^2}{(x^2+y^2)^2}
\end{eqnarray}
with $D = a_3b/4\pi(1-\nu)$ and $\nu = (3a_1-2a_3)/2(3a_1+a_3)$ the Poisson's ratio.

A comparison of the free energy as a function of the dislocation density for these two calculations with a specific values for $c_x$ and $c_y$ for a droplet containing a single grain ($N=1$) of the {\hv II} phase is shown in Fig.\ref{compare}. Taking the magnitude, $b$, of the Burgers vector as an adjustable parameter we see that the optimum for the two alternatives coincide if $b = 1.3 \xi$, an entirely reasonable value. Changing $c_x$ and $c_y$ causes only minor changes in the value of $b$. We therefore identify this solutions as representing dislocated or massive martensite. The correspondence with dislocations also lends meaning to the coefficients $c_x$ and $c_y$ as dislocation mobilities in the glide and climb directions (perpendicular and parallel to the habit plane) respectively, although exact numerical correspondence with measured mobilities for any particular real system is beyond the scope of our theory. Immediately, we note that portions of the dynamical phase diagrams shown in Fig.\ref{2dpd} showing a non-affine phase for $c_y/c_x \gg 1$ may be physically unrealizable for most materials since climb is usually much slower than glide except for very high temperatures where an un-twinned {\em ferrite} microstructure with non-planar interfaces is probably more stable\,\,\citep{paper2}. 

Our theory, incorporating non-affine fields therefore offers a common framework within which both twinned and dislocated martensites in a variety of systems with different dynamical characteristics, each corresponding to physically realizable portions of our dynamical phase diagram Fig.\ref{2dpd}, may be described and discussed. 
\begin{figure}[t]
\begin{center}
\includegraphics[width=14cm]{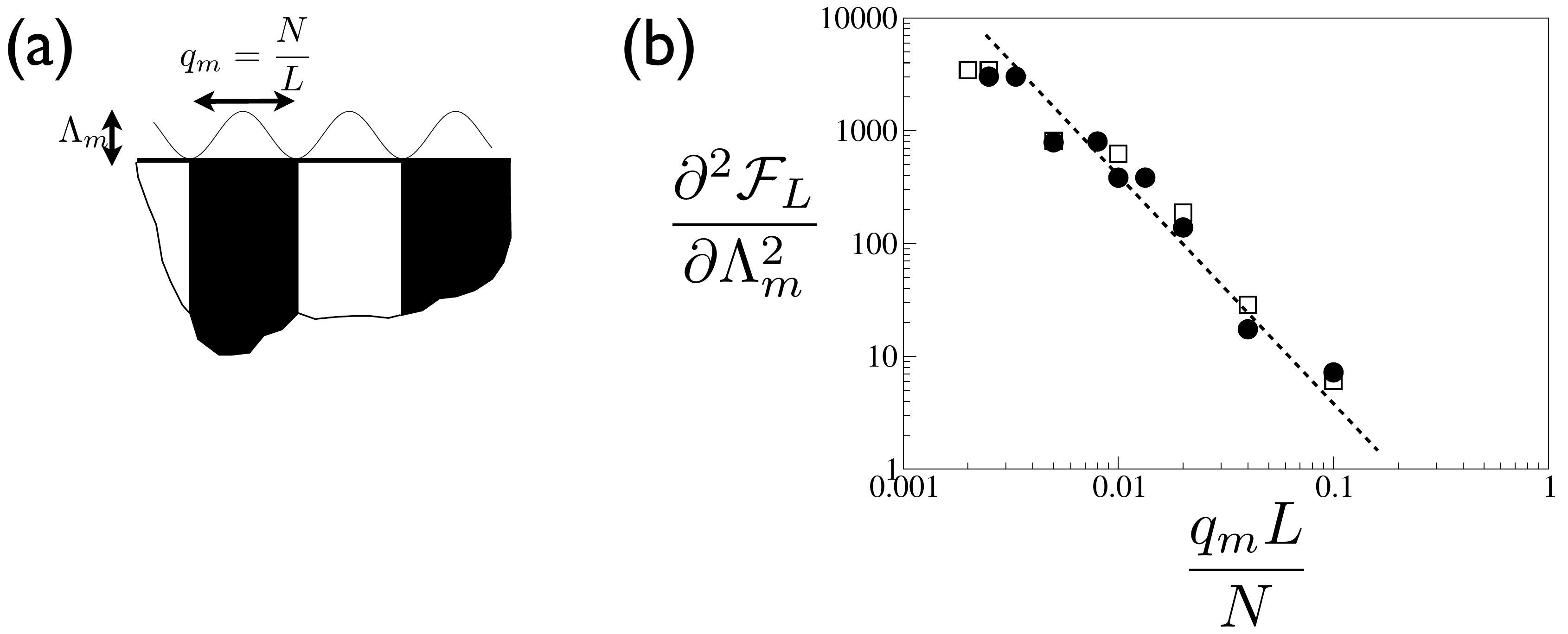}
\end{center}
\caption{(a) Schematic diagram showing the modulation of the habit plane. (b) Plot of the curvature of the Lyapunov free energy ${\cal F}_L$ as a function of the scaled modulation wavenumber for solutions {\hv  II} (filled corcles) and {\hv  III} (open squares). A negative value of this quantity signifies instability. Note that the interface continues to remain planar with the curvature decreasing roughly as $\sim 1/q_m^2$.}
\label{mod}
\end{figure}

\section{Discussion and conclusion}
\label{discussion}
In this paper we have introduced a coarse grained continuum model which extends strain-only theories of martensites by including non-elastic degrees of freedom capable of describing localized regions called NAZ within which particles are known to have non-affine displacements. In the final section of this paper we discuss our present results, taking up a few issues which examine our assumptions, illustrate the relevance of our findings and point out avenues for further work.

\subsection{Stability of the parent-product interface} Our ansatz, for the shape of the martensite droplet is rectangular implying flat twin and parent- product interfaces. How good is this assumption? While twin interfaces are usually flat because they minimize elastic energy, moving parent-product interfaces during growth of the product phase may undergo dynamical (Mullins-Sekerka) instabilities producing a variety of interesting structures and shapes, eg. dendrites \,\citep{mullins}. On the other hand, austenite- martensite interfaces in ferro-elastics remain planar despite the often large velocities involved. This is easy to understand once we realize that the Mullins-Sekerka instability is strongly suppressed at wavenumbers below the inverse capillary length $\propto \Sigma_{eff} \propto \sqrt{W}$ in the limit $W \to \infty$. Since solution {\hv I} corresponds to this case, our ansatz  remains trivially valid. In order to investigate the validity of our ansatz for solutions {\hv  II} and {\hv  III}, we calculate ${\cal F}_L$ numerically after imposing a modulation $\sim \Lambda_m \sin(q_m y)$ (Fig.\ref{mod}(a)) on the habit plane. In Fig.\ref{mod}(b) we show that the curvature of the free energy $\partial^2 {\cal F}_L/\partial \Lambda_m^2 \sim q_m^{-2} > 0$ proving that our ansatz is stable against such modulations. Nevertheless, dynamical instabilities of the Mullins-Sekerka type can still arise since $\phi$ screens elastic interactions leading to a less than $\sqrt{W}$ dependence of the effective surface tension. Indeed austenite- ferrite interfaces which contain many defects \,\citep{mullinsfer} are hardly planar. This has also been observed in our own MD simulations \,\citep{paper1} and elasto-plastic theory \,\citep{paper2}.

\subsection{Stick-slip motion}
Non-affineness at the parent- product interface may produce dynamical phenomena which are qualitatively new.  For small $\phi$, one obtains steady growth of the interface, with $\phi$ being advected along with the growing interface. On the other hand, our threshold dynamics for yield, may give rise to intermittent dissipation into non-affine and affine sectors producing stick-slip dynamics of the austenite-martensite interface which we would like to investigate in the future. Such stick-slip motion has been known to be prevalent in many driven dissipative threshold systems, e.g., frictional sliding \,\citep{fric}, peeling of an adhesive tape \,\citep{adhesive}, earthquake faults \,\citep{earth}, Barkhausen noise in ferromagnetic transitions \,\citep{bark}, serrations in stress-strain curve \,\citep{plc} and rate independent hysteresis \,\citep{Knowles}. The system usually spends a large time in the stuck state during which stresses are build up and subsequent rapid release of stress facilitates the slip of the system. Indeed, even structural transitions in solids, including martensitic transformations \,\citep{martstick,martstick2}, are known to be accompanied by stick-slip motion \,\citep{stickslip}  of the interface. The most predominant view is that stick slip in these systems arises from the presence of compositional disorder \,\citep{kaushik}. However, experiments performed on extremely pure samples having very low compositional disorder \,\citep{arindam} have also exhibited signatures of stick-slip behavior. While pinning by impurities can never be completely eliminated in real systems, one can ask whether such ``external'' pinning is {\em necessary} for stick-slip dynamics in martensites. Can there be intrinsically generated disorder, as represented by our $\phi$ field, which induces stick-slip dynamics during structural transition in pure systems? 

In order to address this question we need to use our phenomenological theory to study the detailed dynamics of the parent~-product interface using the droplet ansatz. Calculations in this direction are underway and will be published elsewhere.

\section{Acknowledgements}
We thank M. Falk, H.K.D.H. Bhadeshia, J. K. Bhattacharya, J. Bhattacharya and T. Das for discussions. We thank the DST, Gov. of India  for support through the Indo-EU project MONAMI and AMRU.

\appendix
\section{Appendix: }
The computation of the contribution $F_{\phi}(\phi_0,\beta_x,\beta_y)$ of the non-affine field $\phi$ to the free energy of a rectangular droplet of size $L$ and $W$ with ${\cal N}$ twins in the limit of $\sigma_{1c} \to 0$ is straight forward, though tedious. The final answer is listed here for reference. 
\begin{eqnarray}
F_{\phi} &=& \frac{1}{2 {\cal N}}\int \left[a_1\phi^2 + 2a_1e_1^A\phi + c_x\left(\frac{\partial \phi}{\partial x}\right)^2 + c_y\left(\frac{\partial \phi}{\partial y}\right)^2\right]dxdy \nonumber \\
&=& \Big[4 \pi a_1\lambda \Big(\frac{L}{N}\Big)^2 +  a_1\lambda h_c(\beta_x,{\cal N})p_c(\beta_y,{\cal N}) \Big] \phi_0 \nonumber \\
& + & \frac{1}{2} \Big[ a_1 h(\beta_x)p(\beta_y,{\cal N})  + c_x h_g(\beta_x,{\cal N})p(\beta_y,{\cal N}) + 
c_y h(\beta_x)p_g(\beta_y,{\cal N}) \Big] \phi_0^2 \nonumber \\
\end{eqnarray}
\noindent
where the functions of $\beta_x$, $\beta_y$ and ${\cal N}$ in the square brackets which are all positive definite are, 
\begin{eqnarray}
h(\beta_x) 
&=& 2\sqrt{\pi}\beta_x\left(1-e^{-W^2/4\beta_x^2}\right),\\
h_c(\beta_x,{\cal N}) 
&=& \sqrt{2\pi}\beta_xe^{q^2\beta_x^2/2}\left[2\left(1-{\rm erf}(q\beta_x/\sqrt{2})\right) \nonumber 
+ e^{-qW}\left({\rm erf}\left(\frac{q\beta_x^2-W}{\sqrt{2}\beta_x}\right)-1\right)\right.\\  
&+&\left. e^{qW}\left({\rm erf} \left(\frac{q\beta_x^2+W}{\sqrt{2}\beta_x}\right)-1\right)\right],\\ 
h_g(\beta_x,{\cal N}) &=& \frac{\sqrt{\pi}}{2\beta_x^3}\left[2\beta_x^2 + \left(W^2-2\beta_x^2\right)e^{-W^2/4\beta_x^2}\right].
\end{eqnarray}
Defining $z=q\beta_y$ and $u=\pi/4z$, we have further,
\begin{eqnarray}
p(\beta_y,{\cal N}) 
&=& \frac{\sqrt{\pi}\beta_y}{4}\left[4e^{-u^2}\big\{{\rm erf}(u) + {\rm erf}(3u)\big\} 
+ 2\big\{{\rm erf}(2u)+{\rm erf}(4u)\big\} + 4e^{-4u^2}{\rm erf}(2u)
\right.  \nonumber \\ 
&+&\left. e^{-z^2}\left\{{\rm erf}(2u+iz)+{\rm erf}(2u-iz)-{\rm erf}(4u+iz)-{\rm erf}(4u-iz)\right\}\right. \nonumber
\\  &+&\left.2ie^{-z^2-u^2}\left\{{\rm erf}(3u+iz)-{\rm erf}(3u-iz)-{\rm erf}(u+iz)+{\rm erf}(u-iz)\right\}\right],\nonumber \\  \\
p_c(\beta_y,{\cal N}) 
&=& \sqrt{\frac{\pi}{8}}\beta_y\left[2\left\{{\rm erf}(\sqrt{2}u) + 
{\rm erf}(2\sqrt{2}u)\right\}\right.  \nonumber \\
&+&\left. e^{-2z^2}\left\{{\rm erf}\left(\sqrt{2}u+i\sqrt{2}z\right) + {\rm erf}\left(\sqrt{2}u-i\sqrt{2}z\right)\right.\right.  \nonumber\\
&-&\left.\left. {\rm erf}\left(2\sqrt{2}u+i\sqrt{2}z\right) - {\rm erf}\left(2\sqrt{2}u-i\sqrt{2}z\right) \right\}\right],  \nonumber \\
p_g(\beta_y,{\cal N}) &=& \frac{\sqrt{\pi}q}{32z^3}e^{-4u^2-z^2}\left[2e^{z^2}\left\{-4\left(\pi^2-2z^2-4z^4\right){\rm erf}(2u)\right.\right. \nonumber \\
&+&\left.\left. e^{3u^2}\left(-\pi^2 + 8z^2 + 16z^4\right)\left({\rm erf}(u) + {\rm erf}(3u)\right)\right.\right.  \nonumber\\ 
&+&\left.\left. 4z^2(1+2z^2)e^{4u^2}\left({\rm erf}(2u) + {\rm erf}(4u)\right)\right\}\right. \nonumber\\
&+&\left.\left(8z^2-4\pi^2+4z^2e^{4u^2}\right)\left({\rm erf}(2u+iz) + {\rm erf}(2u-iz)\right) \right.  \nonumber \\
&-&\left. 4z^2e^{4u^2}\left({\rm erf}(4u+iz) + {\rm erf}(4u-iz)\right)\right.  \nonumber\\
&+&\left. ie^{3u^2}\left({\rm erf}(u+iz)-{\rm erf}(u-iz)\right) \right.  \nonumber\\
&-&\left. i(\pi^2-8z^2)e^{3u^2}\left({\rm erf}(3u+iz)-{\rm erf}(3u-iz)\right)\right]
\end{eqnarray}

\bibliographystyle{plainnat}
\bibliography{references}

\end{document}